\documentclass[]{aastex631}

\usepackage{amsmath}

\shorttitle{Analytical GI wiggle}
\shortauthors{Longarini et al.}
\graphicspath{{./}{figures/}}
\begin{document}

\title{Investigating protoplanetary disc cooling through kinematics: analytical GI wiggle}

\correspondingauthor{Cristiano Longarini}
\email{cristiano.longarini@unimi.it}

\author[0000-0003-4663-0318]{Cristiano Longarini}
\affiliation{Dipartimento di Fisica, Università degli Studi di Milano, via Celoria 16, 20133 Milano, Italy}

\author[0000-0002-2357-7692]{Giuseppe Lodato}
\affiliation{Dipartimento di Fisica, Università degli Studi di Milano, via Celoria 16, 20133 Milano, Italy}

\author[0000-0002-6958-4986]{Claudia Toci}
\affiliation{Dipartimento di Fisica, Università degli Studi di Milano, via Celoria 16, 20133 Milano, Italy}

\author[0000-0002-2611-7931]{Benedetta Veronesi}
\affiliation{Dipartimento di Fisica, Università degli Studi di Milano, via Celoria 16, 20133 Milano, Italy}
\affiliation{Univ Lyon, Univ Lyon1, Ens de Lyon, CNRS, Centre de Recherche Astrophysique de Lyon UMR5574, F-69230, Saint-Genis,-Laval, France.}

\author[0000-0002-8138-0425]{Cassandra Hall}
\affiliation{Department of Physics and Astronomy, The University of Georgia, Athens, GA 30602, USA}
\affiliation{Center for Simulational Physics, The University of Georgia, Athens, GA 30602, USA}

\author[0000-0001-9290-7846]{Ruobing Dong}
\affiliation{Department of Physics \& Astronomy, University of Victoria, 3800 Finnerty Road, Victoria, BC V8P 5C2, Canada}

\author[0000-0002-8590-7271]{Jason Patrick Terry}
\affiliation{Department of Physics and Astronomy, The University of Georgia, Athens, GA 30602, USA}
\affiliation{Center for Simulational Physics, The University of Georgia, Athens, GA 30602, USA}

%% Note that the \and command from previous versions of AASTeX is now
%% depreciated in this version as it is no longer necessary. AASTeX 
%% automatically takes care of all commas and "and"s between authors names.

%% AASTeX 6.31 has the new \collaboration and \nocollaboration commands to
%% provide the collaboration status of a group of authors. These commands 
%% can be used either before or after the list of corresponding authors. The
%% argument for \collaboration is the collaboration identifier. Authors are
%% encouraged to surround collaboration identifiers with ()s. The 
%% \nocollaboration command takes no argument and exists to indicate that
%% the nearby authors are not part of surrounding collaborations.
\defcitealias{Hall2020}{H20}
%% Mark off the abstract in the ``abstract'' environment. 
\begin{abstract}
It is likely that young protostellar discs undergo a self-gravitating phase. Such systems are characterised by the presence of a spiral pattern that can be either in a quasi-steady state or in a non-linear unstable condition. This spiral wave affects both the gas dynamics and kinematics, resulting in deviations from the Keplerian rotation. Recently, a lot of attention has been devoted to  kinematic studies of planet forming environments, and we are now able to measure even small perturbations of velocity field  {($\lesssim 1\%$ of the Keplerian speed)} thanks to high spatial and spectral resolution observations of protostellar discs. In this work, we investigate the kinematic signatures of gravitational instability: we perform an analytical study of the linear response of a self-gravitating disc to a spiral-like perturbation, focusing our attention on the velocity field perturbations. We show that  unstable discs have clear kinematic imprints into the gas component across the entire disc extent, due to the GI spiral wave perturbation, resulting in deviations from Keplerian rotation. The shape of these signatures depends on several parameters, but they are significantly affected by the cooling factor: by detecting these features, we can put constraints on protoplanetary discs cooling. 
\end{abstract}

%% Keywords should appear after the \end{abstract} command. 
%% The AAS Journals now uses Unified Astronomy Thesaurus concepts:
%% https://astrothesaurus.org
%% You will be asked to selected these concepts during the submission process
%% but this old "keyword" functionality is maintained in case authors want
%% to include these concepts in their preprints.

%% From the front matter, we move on to the body of the paper.
%% Sections are demarcated by \section and \subsection, respectively.
%% Observe the use of the LaTeX \label
%% command after the \subsection to give a symbolic KEY to the
%% subsection for cross-referencing in a \ref command.
%% You can use LaTeX's \ref and \label commands to keep track of
%% cross-references to sections, equations, tables, and figures.
%% That way, if you change the order of any elements, LaTeX will
%% automatically renumber them.
%%
%% We recommend that authors also use the natbib \citep
%% and \citet commands to identify citations.  The citations are
%% tied to the reference list via symbolic KEYs. The KEY corresponds
%% to the KEY in the \bibitem in the reference list below. 
\section{Introduction}
For cold massive discs, the role of the disc self-gravity becomes dynamically important, affecting the vertical and radial structure of the system \citep{KrattLod,BerLod}. In this context, gravitational instabilities (hereafter referred as GI) often arise, determining the evolution of the system and playing a fundamental role in the transport of angular momentum. The development of the GI has initially been studied in the context of galactic dynamics \citep{LinShu,BertinLin,BinneyTremaine}: as far as protostellar discs are concerned, the results are quantitatively similar.
%( In the last years, crucial efforts were put into the study of pseudo-viscous models of self-gravitating discs \citep{clarkepseudo,ricearm}, providing an equivalent $\alpha$-prescription.)

On the one hand, a possible outcome of GI in protostellar environments is the fragmentation of the disc: this phenomenon can potentially lead to the formation of low-mass stellar companions \citep{frag1,frag2,frag3}, because the initial clump mass is of the order of several Jupiter masses, too high to form a planet \citep{KrattLod}. On the other hand, GI is a very effective way to transport angular momentum within the disc, by means of a global spiral perturbation \citep{transport1,transport2}. 

High resolution observations with ALMA have revealed that most of the observed protostellar discs possess substructures as rings or spirals. The origin of rings is often explained by planets \citep{planet1,dipierroelias,planet2,planet3}, however what causes the spirals is still ambiguous. Indeed, super-Jupiter objects can excite spiral density waves with azimuthal wavenumber $m\sim1-2$ that match with good agreement the observed structures in scattered light \citep{dongscatt,dongfung,benniSAO,rosottisp}. In addition, some spirals may also be induced by a inner or outer stellar companion \citep{companion}, or by a flyby \citep{flyby1,flyby2}. At the same time, large scale spiral perturbations also characterise self-gravitating discs, with a typical $m\sim M_\star/M_d$, where $M_d$ is the disc mass and $M_\star$ is the mass of the star \citep{Cossins}. Distinguishing the origin of a spiral is difficult, but recent high resolution observations of protostellar environments allow us to conduct kinematic studies that might shed some light on this issue. It is well known that the presence of a perturber inside the disc creates a localised deviation from the Keplerian observed velocity, called ``kink'' \citep{kink1,kink2}: when the perturber is a planet, the kink can be used as a proxy for its mass \citep{planetkinks}. As far as GI is concerned, \citet{Hall2020} (hereafter referred as \citetalias{Hall2020}) show, based on hydrodynamical simulations, that the spiral perturbation deeply affects the gas kinematic: in particular, it creates a global (rather than a localised) deviation from Keplerian observed velocity - a ``global kink'' - dubbed GI wiggle by \citetalias{Hall2020}, that is apparent in the moment one and in the channel maps \citep{teresa}.

In this work, we present an analytical study of the response of a self-gravitating protostellar disc to a spiral density wave in WKB regime. We focus our attention on the velocity field perturbations (hereafter referred as VPs), and we show that they can be written as a function of disc parameters. Thanks to such analytical expressions, we are able to sketch the observed velocity field, and then to make a connection to observations. The amplitude of the VPs is linked to the cooling factor of the system, and thus we can use this relation to investigate the cooling in protostellar discs. 

The paper is organised as follows. In section \ref{S2} we summarise the theory that we need to conduct our study, paying attention to both  dynamical and thermodynamical processes. In section \ref{S3} we obtain the analytical expression of the VPs in WKB regime. In section \ref{S4} we discuss the observational aspects of the wiggles, projecting the perturbed velocity field along the line of sight. To conclude, in section \ref{S5} we discuss the shape of the perturbations, the limitations of this work and we present a mock experiment to test our predictions.

\section{Self-gravitating gaseous discs}\label{S2}
\subsection{The disc potential}
The gravitational potential of a self-gravitating protostellar disc is not exactly Keplerian: indeed, when $M_d \gtrsim (H/r) M_\star$, where $H/r$ is the aspect ratio of the disc, the disc contribution to the potential is not negligible. In this case, the radial balance of forces for a cold disc is given by 
\begin{equation}\label{omega}
    \Omega^{2} =  \frac{G M_{\star}}{r^{3}} + \frac{1}{r} \frac{\partial \Phi_{d}}{\partial r},
\end{equation}
where the first term is the Keplerian frequency and the second one is the disc contribution \citep{BerLod}. The gravitational field generated by the disc can be written as
\begin{equation}\label{discvel}
    \frac{\partial \Phi_{d}}{\partial r}(r, z)=\frac{G}{r} \int_{0}^{\infty} d r^{\prime}\left[K(\zeta)-\frac{1}{4}\left(\frac{\zeta^{2}}{1-\zeta^{2}}\right)  \times\left(\frac{r^{\prime}}{r}-\frac{r}{r^{\prime}}+\frac{z^{2}}{r r^{\prime}}\right) E(\zeta)\right] \sqrt{\frac{r^{\prime}}{r}} \zeta \Sigma\left(r^{\prime}\right),
\end{equation}
where $E(\zeta),$ $K(\zeta)$ are the complete elliptic integrals of the first kind, $\zeta^2 =4rr^{\prime}/[(r+r^\prime)^2+z^2]$ and $\Sigma$ is the disc surface density. Deviations from the Keplerian behaviour have actually been seen in the rotation curve of Elias 2-27 \citep{bennielias}, and this can be an effective method for measuring the disc mass. 

\subsection{Gravitational instability}
%The development of the gravitational instability in self-gravitating discs has initially been studied in the context of galactic dynamics \citep{LinShu,BertinLin,BinneyTremaine}: as far as protostellar discs are concerned, the results are quantitatively similar \citep{KrattLod}.
The linear response to gravitational instability is described by the well known dispersion relation \citep{LinShu}
\begin{equation}\label{disp}
    D(\omega, k, m)=(\omega-m \Omega)^{2}-c_{\mathrm{s}}^{2} k^{2}+2 \pi G \Sigma|k|-\kappa^{2}=0,
\end{equation}
where $\omega$ is the wave angular frequency, $k$ the radial wavenumber, $m$ the azimuthal wavenumber, $\kappa(r)$ the epicyclic frequency, $\Omega(r)$ the angular frequency and $c_s(r)$ the sound speed. This relationship has been obtained for an infinitesimally thin disc under the WKB perturbation analysis and in the tight winding limit (i.e. the radial wavelength is much smaller than the azimuthal one). The stability criterion can be expressed by means of the $Q$ parameter
\begin{equation}
    Q = \frac{c_s \kappa}{\pi G \Sigma},
\end{equation}
that contains the stabilising terms at numerator and the unstable ones at denominator. The instability threshold is given by $Q=1$: if $Q>1$ the disc is stable at all wavelengths while $Q<1$ identifies a range of unstable wavelengths. For the case of an unstable disc, the most unstable wavenumber is 
\begin{equation}
    k_\text{uns} = H_\text{sg}^{-1} = \frac{\pi G \Sigma}{c_s^2},
\end{equation}
where $H_\text{sg} = c_s^2 / \pi G \Sigma$ is the disc thickness in the self gravitating case.

A marginally stable disc is characterised by having $Q\simeq1$: in this case, the only expected excited modes would be $k\simeq k_\text{uns}$. Thus equation (\ref{disp}) tells that $\Omega\simeq\omega/m=\Omega_\text{p}$, where $\Omega_\text{p}$ is the spiral pattern frequency, meaning that all excited modes are expected to be close to corotation \citep{Cossins}.

\subsubsection{Spiral density waves}
A spiral wave is characterised by having
\begin{equation}\label{mod}
    m\phi + \psi(r) = \text{const}, \quad \text{mod } 2\pi,
\end{equation}
where $\psi(r)$ is the shape function and it holds that $k = \text{d}\psi/\text{d}r$. It is useful to introduce the radial wavenumber $k$, that is the radial derivative of the shape function; the sign on $k$ determines wheter the spiral wave is leading ($k<0$) or trailing ($k>0$). Another important quantity is the opening angle of the spiral $\alpha_p$, hereafter referred as the pitch angle: it is given by 
\begin{equation}
    \tan\alpha_p = \left|r\frac{\partial \phi}{\partial r}\right|^{-1},
\end{equation}
where the partial derivative is evaluated along (\ref{mod}), giving $\tan\alpha_p = m/rk$.

If we consider a self-consistent spiral perturbation, we can easily link the density perturbation to the potential one \citep{Cossins}. The perturbed surface density can be written as 
\begin{equation}
    \Sigma_1(r,\phi,t) = \text{Re}\left[\delta \Sigma(r) e^{i(m\phi-\omega t + \psi(r))}\right],
\end{equation}
and it can be shown  that the corresponding perturbed potential is given by
\begin{equation}\label{hp2}
    \Phi_1(r,\phi,t)=-\frac{2\pi G}{|k|}\Sigma_1(r,\phi,t).
\end{equation}

\subsubsection{Thermodynamics}
So far, we have discussed only the linear growth of gravitational instability, however, to properly understand the outcome of this process, we need to consider the non-linear evolution. To do so, it is necessary to introduce a parameter to capture
the radiative processes of the disc. 

The non-linear growth of perturbations is best understood by using numerical simulations \citep{nonlin1,nonlin2,nonlin3}. However, reproducing realistically the cooling processes is not an easy task: in the last years, a lot of effort was devoted to the  {modeling of realistic thermodynamics} \citep{cool1,cool2,hiroseshi}.

In this paper, we are not interested in the physics of cooling, but in the relationship between the density perturbations and the rate at which the disc cools. For this reason, we impose a  {prescribed cooling law}
\begin{equation}
    q^{-} = -\frac{e}{t_\text{cool}},
\end{equation}
where $e$ is the internal energy per unit mass and all details of the cooling are absorbed by $t_\text{cool}$. This parameter defines a typical timescale, regardless of what process we are taking into account. Often, the ratio between the cooling time and the dynamical one is chosen to be constant, such that $\beta = \Omega t_\text{cool}=\text{const}$  \citep{gammie01}. In this work, we use this prescription and hereafter we will refer to the cooling process in terms of $\beta$.

The stability parameter $Q$ is proportional to the sound speed, thus to the temperature, so that colder discs are prone to being unstable. In absence of external heating mechanisms, an initially stable hot disc $(Q>>1)$ will cool down due to radiative processes, until eventually reaching the marginally stable state $Q\simeq 1$. At this point, gravitational instability turns on: the disc develops a spiral structure that, by means of compression and shocks, leads to an efficient energy dissipation and heating. In this sense, the $Q-$stability condition acts as a thermostat so that
heating turns on only if the system is sufficiently cold, keeping it in a marginal stable state $(Q\simeq 1)$. In this regime, spiral perturbations do not grow exponentially, but their amplitude saturates at some finite value. In order to establish the conditions under which the disc self regulates, we need to take into account the heating processes too. As we have said, the generation of spiral density waves leads to propagation of shocks, because of the supersonic difference of speed between the spiral pattern and the underneath disc. The self regulation condition can be obtained by balancing cooling and heating terms \citep{Cossins}, and it gives 
\begin{equation}\label{cool}
    \frac{\delta \Sigma}{\Sigma}=\chi \beta^{-1/2},
\end{equation}
where $\chi$ is the proportionality factor, that is of the order of unity, and $\beta$ is the adimensional cooling factor. Hence, a more efficient cooling (lower $\beta$) gives rise to stronger density perturbations. Note that the amplitude of the density perturbations is non-linear: indeed, what we did here is to take into account non-linearities to relate the expected amplitude to the cooling rate.

 {We should remember that equation (\ref{cool}) has been obtained under the hypothesis that the only heating process in the system is the propagation of shocks, induced by spiral density waves. If we consider irradiated disks, i.e. disks that are also heated by the central star, the self regulation condition is different. In particular, irradiation is known to reduce the amplitude of SG perturbations: \citet{irrdisks} studied the problem of fragmentation in irradiated disks, and they found that the fragmentation threshold could decline by approximately a factor of two. Thus, by neglecting this effect, we expect that in an actual system for given $\delta\Sigma/\Sigma$, the $\beta$ factor should be overestimated.}

\section{Velocity perturbations}\label{S3}
In this section, we start from the fluid equations for an infinitesimally thin self-gravitating accretion disc, we perturb them with a spiral-like disturbance and we extract the velocity perturbations, following the formalism of \citet{BinneyTremaine}. We use cylindrical coordinates $(r,\phi,z)$ and we restrict to the $z=0$ plane. The dynamics is characterised by the continuity equation (equation \ref{cont}), the two components of Euler's equation (equations \ref{eul}), the Poisson's equation (equations \ref{poi}) and the equation of state of the fluid (equation \ref{enth})
\begin{equation}\label{cont}
    \partial_t \Sigma + \frac{1}{r} \partial_r (\Sigma u_r r) + \frac{1}{r} \partial_\phi(\Sigma u_\phi) =0,
\end{equation}
\begin{subequations}\label{eul}
\begin{align}
      & \partial_t u_r + u_r \partial_r u_r + \frac{u_\phi}{r} \partial_\phi u_r -\frac{u_\phi ^2}{r} =  -  \partial_r (\Phi +h),\\
      & \partial_t u_\phi + u_r  \partial_r u_\phi + \frac{u_\phi}{r} \partial_\phi u_\phi + \frac{u_\phi u_r}{r} = -\frac{1}{r}\partial_\phi (\Phi +h),
\end{align}
\end{subequations}
\begin{equation}\label{poi}
    \nabla^2 \Phi = 4\pi G \Sigma \delta(z),
\end{equation}
\begin{equation}\label{enth}
    dh = c_s ^2 \frac{d\Sigma}{\Sigma},
\end{equation}
where $h$ is the enthalpy of the fluid, $u_r$ and $u_\phi$ the radial and azimuthal component of the velocity field.

For linear analysis, we assume that the spiral perturbation is small compared to the disc background, and hence can be Fourier-decomposed in time $t$ and azimuthal angle $\phi$. All the variables $(\Sigma,u_\phi,u_r,h,\Phi)$ can be written as $X = X_0(r) + X_1(r,\phi,t)$, where $X_0$ refers to the basic state and $X_1$ to the perturbation. The basic state of the system is given by $\Sigma_0$, $u_{\phi0}=r\Omega$, where $\Omega$ is given by equation (\ref{omega}), $u_{r0}=0$, $h_0$.

Keeping only terms that are first order in $X_1$, we get the well known equations for the velocity perturbations
\begin{equation}\begin{split}
    &  u_{r1} = \frac{i}{\Delta} \left[(\omega - m \Omega)\partial_r(\Phi_1 + h_1) - \frac{2m\Omega}{r} (\Phi_1 + h_1)\right],\\
    &  u_{\phi1} = -\frac{1}{\Delta}\left[2B\partial_r( \Phi_1 +  h_1) + \frac{m(\omega - m \Omega)}{r} ( \Phi_1 +  h_1)\right],
\end{split}\end{equation}
where $B(r) = -\frac{1}{2} \frac{\text{d}(\Omega r)}{\text{d}r} + \Omega$ is one of the Oort parameter \citep{oort}, $\Phi_1$ is given by (\ref{hp2}), $  h_1 = c_s^2 {\Sigma_1}/{\Sigma_0}$ and $\Delta = \kappa^2 -(\omega -m\Omega)^2$.

Now we make some assumptions: firstly, we write the perturbation as
\begin{equation}
    X_1 = \text{Re}\left[\delta X(r)e^{i(m\phi-\omega t + \psi)}  \right], 
\end{equation}
where $\delta X$ is exclusively a function of the radius. Secondly, we consider a marginally stable accretion disc with $Q=1$, meaning that $\Delta = \kappa^2$ and $k = k_\text{uns}$. Thirdly, both the potential ($\delta\Phi$) and the enthalpy ($\delta h$) perturbations are linked to the density one that, for a self regulate state, is connected to the basic state through the cooling rate (\ref{cool}). Hence, we have found a way to express the perturbations as a function of the basic quantities and the cooling $\beta$:
\begin{equation}
    \delta\Phi = -\frac{2\pi G }{|k|} \delta\Sigma = -2c_s^2 \chi \beta^{-1/2},
\end{equation}
\begin{equation}
    \delta h = c_s^2 \frac{\delta \Sigma}{\Sigma_0} = c_s^2\chi\beta^{-1/2} = -\frac{1}{2}\delta\Phi.
\end{equation}
Finally, the velocity perturbations become 
\begin{equation}\label{vPs}
\begin{split}
   & \delta u_{r} = \frac{2im\Omega\chi}{r\kappa^2}\beta^{-1/2} c_s^2,\\
   & \delta u_{\phi} = \frac{2iB\chi}{\kappa^2}  \frac{\text{d}\psi}{\text{d}r}\beta^{-1/2}c_s^2.
   \end{split}
\end{equation}

\subsection{Nearly Keplerian disc}
In this paragraph, we write the velocity perturbations for a nearly Keplerian disc. This regime is identified by the conditions that 
\begin{equation}
    \frac{\kappa - \Omega_k}{\Omega_k} <1, \quad \frac{\Omega -\Omega_k}{\Omega_k} <1,
\end{equation}
where with the subscript $k$ we identify the Keplerian quantities. This assumption allows us to write $\kappa\simeq\Omega\simeq\Omega_k\propto {r}^{-3/2}$ and $B\simeq-\Omega/4$. With these assumptions, the equations for the VPs are
\begin{equation}\label{vpsdef}
\begin{split}
    & \delta u_{r} = 2im\chi\beta^{-1/2}\left(\frac{M_d(r)}{M_\star} \right)^2u_k \\
    & \delta u_{\phi} = -\frac{i\chi\beta^{-1/2}}{2}\left(\frac{M_d(r)}{M_\star} \right) u_k,
\end{split}
\end{equation}
 {where $u_k$ is the Keplerian speed}, and then the velocity field is given by $u_r = \text{Re}\left[\delta u_r e^{i(m\phi-\omega t + \psi)}\right]$ and $u_\phi = r\Omega + \text{Re}\left[\delta u_\phi e^{i(m\phi-\omega t + \psi)} \right]$, where $\Omega$ is given by equation (\ref{omega}). To obtain these expressions we used that $Q=1$ and that $\text{d}\psi / \text{d}r = k = k_\text{uns}$. Note that $\delta u_{r}/\delta u_{\phi} \simeq 4m M_d/M_\star$  {: for example, when $m=2$, a relatively light disk having $M_d = 0.125 M_\star$ has $\delta u_r = \delta u_\phi$.}

 {In the analysis above, we have neglected the effect of pressure gradients. This is for two main reasons: firstly, it influences only the basic state of the system, not the perturbations, at leat to first order. Secondly, when we consider a marginally unstable self-gravitating disk ($Q=1$), we expect the contribution of the pressure gradient to the velocity field to be sub-dominant with respect to the self-gravity one. Indeed, for such a disk, the self-gravitating contribution is of the order of $H/r$, while the pressure term is $O(H^2/r^2)$ \citep{KrattLod,bennielias}. The effects of pressure gradients are stronger when considering much lower disk/star mass ratios (e.g., see \citealt{rosen}). For the massive disks that we consider in this work, the pressure gradient can thus be neglected. In the light of this, while considering the pressure gradient is critical when one wants to explore the basic state, as done in \citet{bennielias}, this is not strictly necessary in our perturbation theory.}

\subsection{Not constant cooling factor}\label{varying}
So far we have considered only the case of constant $\beta-$cooling. In principle, however, self consistent models of GI discs \citep{clarkepseudo,ricearm} show that $\beta$ varies with the radius \citep{hallspiral}. This happens because these models give a realistic cooling prescription, i.e. radiative cooling, and its rate depends on the temperature of the disc at the mid-plane and on the Rosseland opacity \citep{bellin}. If one sets the density profile to be a power law with radius, the cooling prescription can be written as a collection of power laws with indices $n_i$, depending on the density and the temperature. 

In general, we can chose any cooling law $\beta(r, \rho(r),T(r))$ and then obtain the VPs thorugh equations (\ref{vPs}).

\section{Connection with observations: moment one and channel maps}\label{S4}
In the previous section, we have computed the velocity perturbations due to the presence of a spiral density wave. Now, we want to connect what we have found to observations: what are the observational imprints of these perturbations? The observed velocity field of the gas is obtained by calculating the intensity weighted average velocity of the emission line profile, i.e. the ``moment-1" map. In this work, we do not take into account line emission processes, and instead we simply compute the projected velocity field along the line of sight and we study the impact of velocity perturbations that we have just obtained.

We assume two dimensional polar system of coordinates $(r,\phi)$ centred upon the the star, so that the velocity vector can be written as $\mathbf{u}=(u_r,u_\phi)$. We consider the disc inclined with an angle $\theta$, that in the following we take equal to $\pi/6$. Within this framework, the projected velocity field can be written as
\begin{equation}
    v_\text{obs} = u_r\sin\phi\sin\theta + u_\phi\cos\phi\sin\theta + v_\text{syst},
\end{equation}
where $v_\text{syst}$ is the velocity of the system projected towards the line of sight. A channel map is defined as the isovelocity contours for a chosen observed velocity. For a purely Keplerian disc, $u_r=0$ and $u_\phi = u_k = \sqrt{GM_\star/r}$, and we obtain the well known ``butterfly pattern".% We want to underline that in this work we consider the disc to be face on, and the observer inclined with an angle $\theta$. This allows us to appreciate better the characteristic of the projected velocity field. 

%\begin{figure}
%\gridline{\fig{images/sketch.001.png}{0.6\textwidth}{}}
   % \caption{Geometry of the system disc-observer. We recall that $\theta$ is the inclination and $\phi$ is the azimuthal angle. In this work, we consider the disc to be face on and the observer inclined with an angle $\theta$.}
  %  \label{geometry}
%\end{figure}

\begin{figure*}
\gridline{\fig{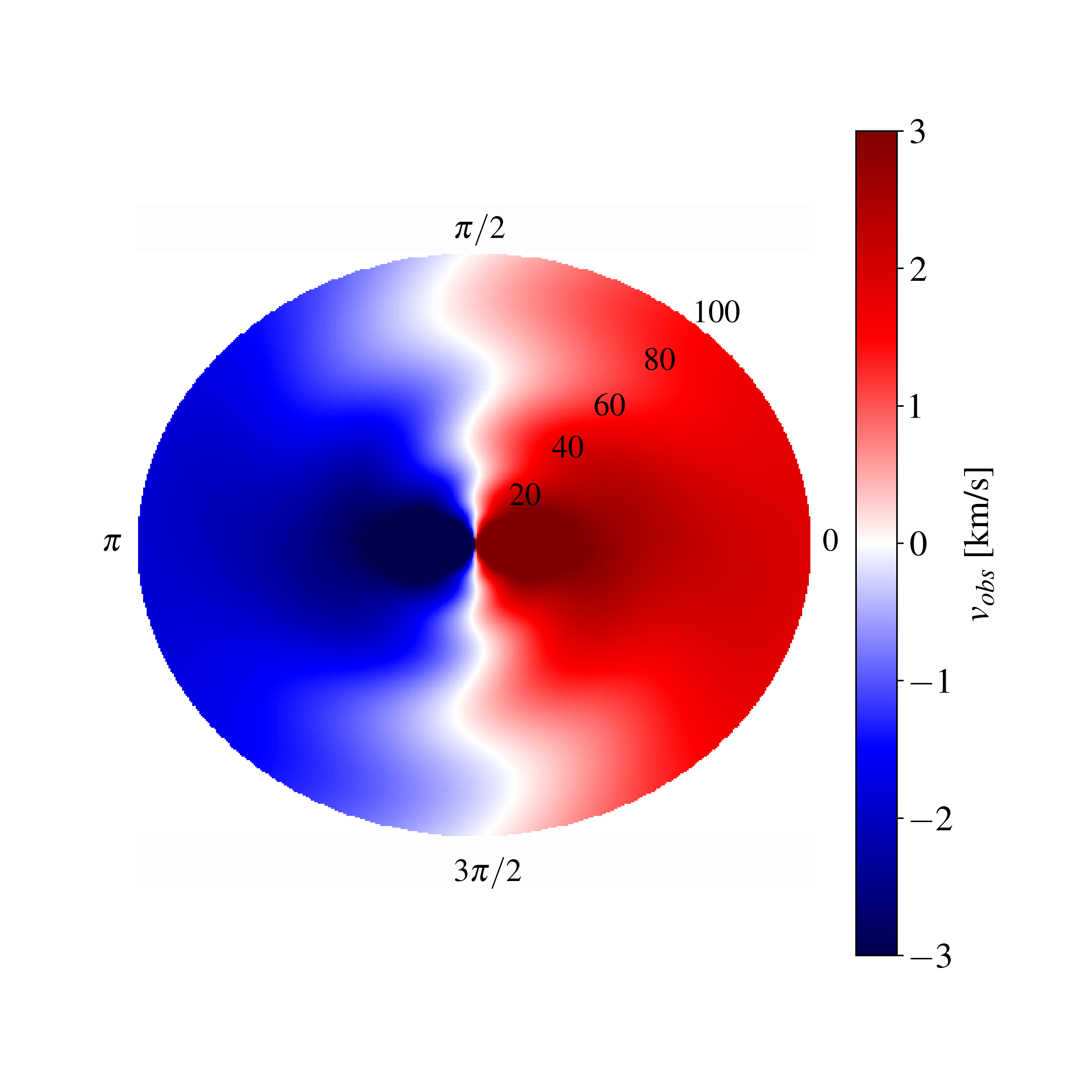}{0.45\textwidth}{(a)}
          }
\gridline{\fig{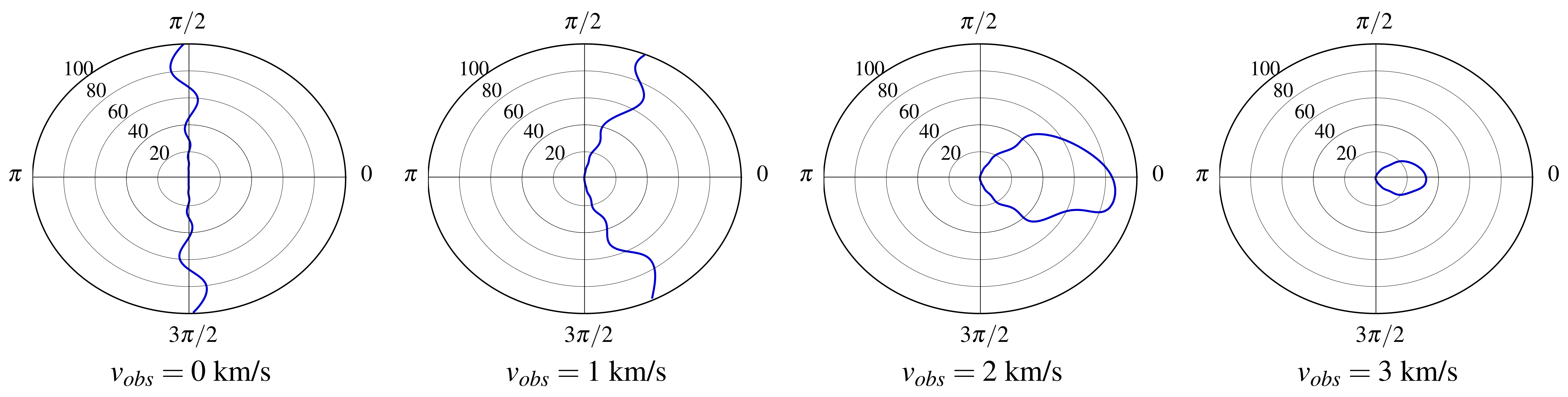}{0.9\textwidth}{(b)}
          }
\gridline{\fig{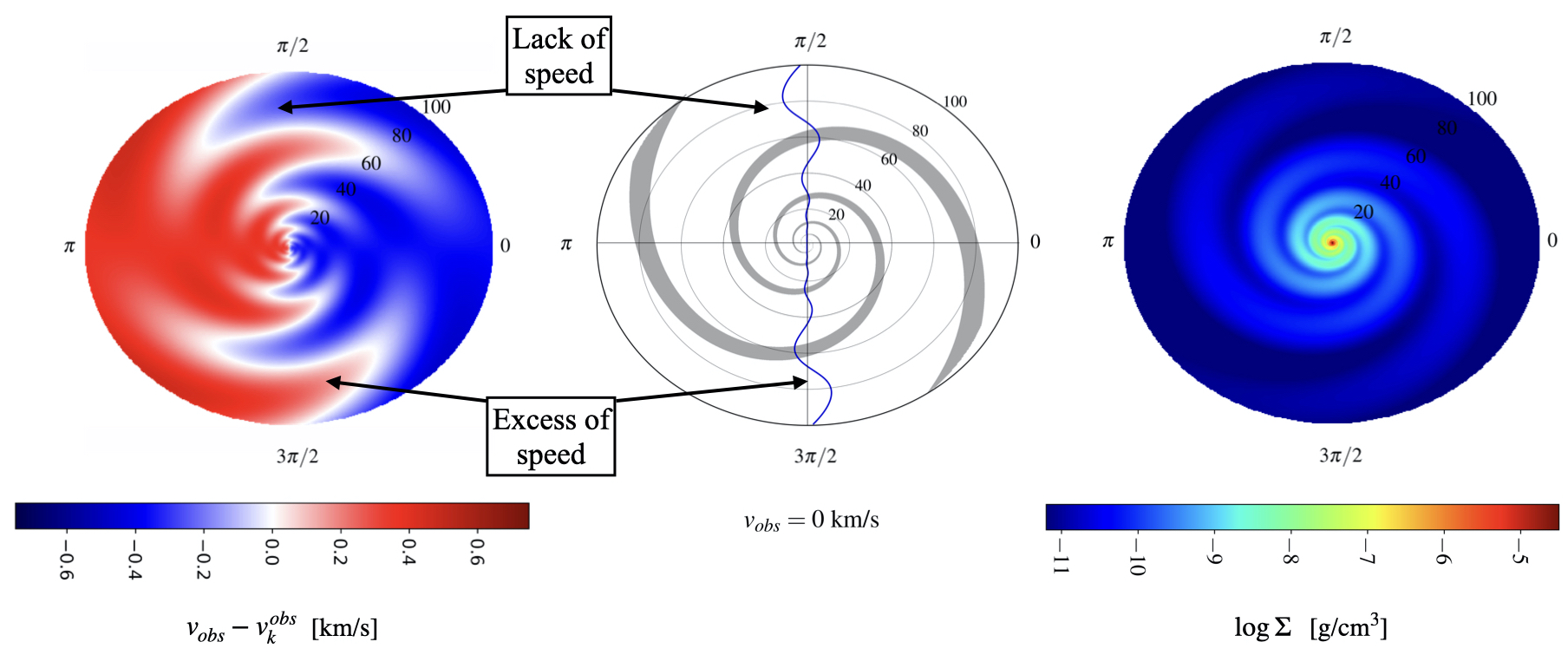}{0.81\textwidth}{(c)}
            }
\caption{Moment one map (a) and channel maps (b) for a self gravitating accretion disc seen with an inclination angle of $\pi/6$ and with a systemic velocity $v_\text{syst} = 0$. (c) Left panel: projected map of the velocity perturbation, after subtraction of the Keplerian field. A system of interlocking fingers is clearly visible, as already noted by \citetalias{Hall2020}. Central panel: the $v_\text{obs}=0$ contour (blue line) overlaid with the spiral shape (grey line). The deviations from the Keplerian channel (that is simply a straight line) perfectly match with the spiral pattern. Right panel: surface density of the disc. The parameters of the disc are the following: $r_\text{in}={1}\text{au}$, $r_\text{out}={100 }\text{au}$, $M_\star = {1}\text{M}_{\odot}$, $M_d = 0.3\text{M}_\star$, $p=-1$, $\beta = 5$, $\alpha_p = 15^\circ$ and $m=2$.}
\label{M1ch}
\end{figure*}

Hence, once we know the radial and the azimuthal components of the perturbed velocity field (equations \ref{vpsdef}), we can sketch the projected velocity field (moment-1 equivalent) and the channel maps, as shown in panels (a), (b) of figure \ref{M1ch}. As already noted in \citetalias{Hall2020}, the VPs due to gravitational instability appear throughout the whole extent of the disc, rather than being localised in position and velocity, as occurs for the kink produced by an embedded protoplanet. This is clearly shown in panel (c) of figure \ref {M1ch}, where we subtract the Keplerian field to the perturbed one: an ``interlocking fingers" structure is present, as already pointed out in \citetalias{Hall2020}. If we look at the central channel, the deviations from the Keplerian behaviour exactly match with the fingers pattern.

We want to underline that in this work we are only considering the projection of the velocity field along the line of sight, without making any assumptions about the gas emission processes. In order to convert velocities to fluxes, it is necessary to include the physics of the gas, specifying the selected tracer and the emission lines observed and considering also the effect of the beam size and the eventual presence of observational noise. To do so, we should use radiative transfer codes, as done in \citetalias{Hall2020}.

\section{Discussion and Conclusions}\label{S5}
\subsection{Shape of the wiggle}
\begin{figure*}
	\gridline{\fig{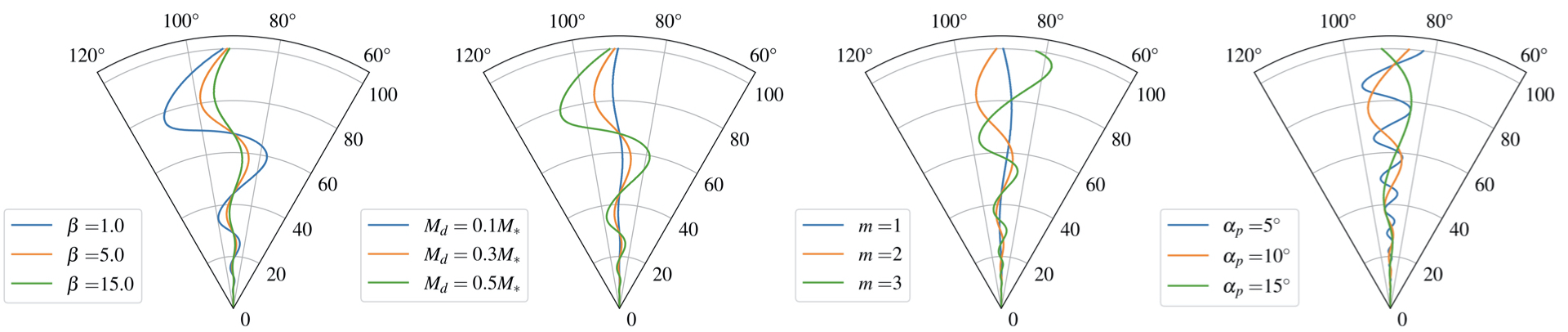}{0.99\linewidth}{}}
    \caption{Shape of the wiggle varying (from left to right) the cooling factor $\beta$, the mass of the disc $M_d$, the pitch angle $\alpha_p$ in degrees and the azimuthal wavenumber $m$. The reference disc parameters are $r_\text{in}={1}\text{au}$, $r_\text{out}={50}\text{au}$, $M_\star = {1}{\text{M}_\odot}$, $M_d = 0.3\text{M}_\star$, $p=-1$, $\beta = 5$, $\alpha_p = 15^\circ$ and $m=2$.}
    \label{wiggA}
\end{figure*}

So far, we have seen that deviations from the Keplerian behaviour are visible in every channel map, however we now focus on the central velocity channel that, in a case where $v_\text{syst}=0$, corresponds to  $v_\text{obs}=0$. As far as the Keplerian case is concerned, the central channel is simply a straight line, because the radial velocity is zero. In the spiral-perturbed case, the central channel shows oscillation around the Keplerian behaviour (i.e. the wiggle): this happens because the spiral wave perturbs both the azimuthal and the radial component. The amplitude and the radial frequency of the wiggle depend on the strength of the spiral wave (that is related to the cooling factor $\beta$ ), the opening angle $\alpha_p$ and the number of arms $m$ (figure \ref{wiggA}), and on the structure of the underlying disc, in particular its mass.

\begin{figure}
\gridline{\fig{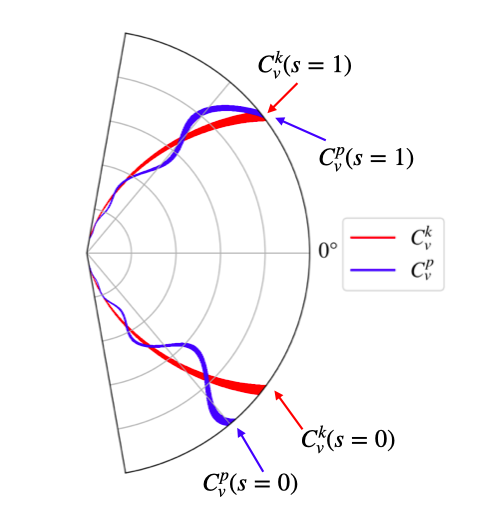}{0.32\textwidth}{(a)}
\fig{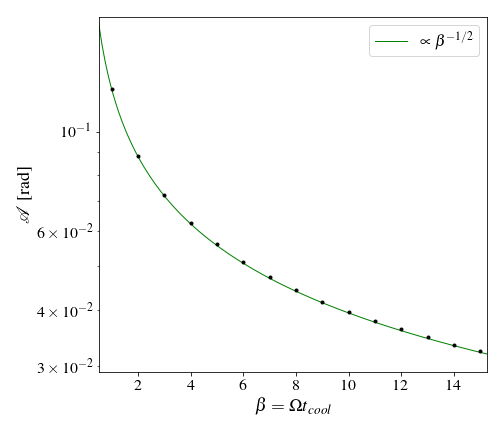}{0.325\textwidth}{(b)}   
\fig{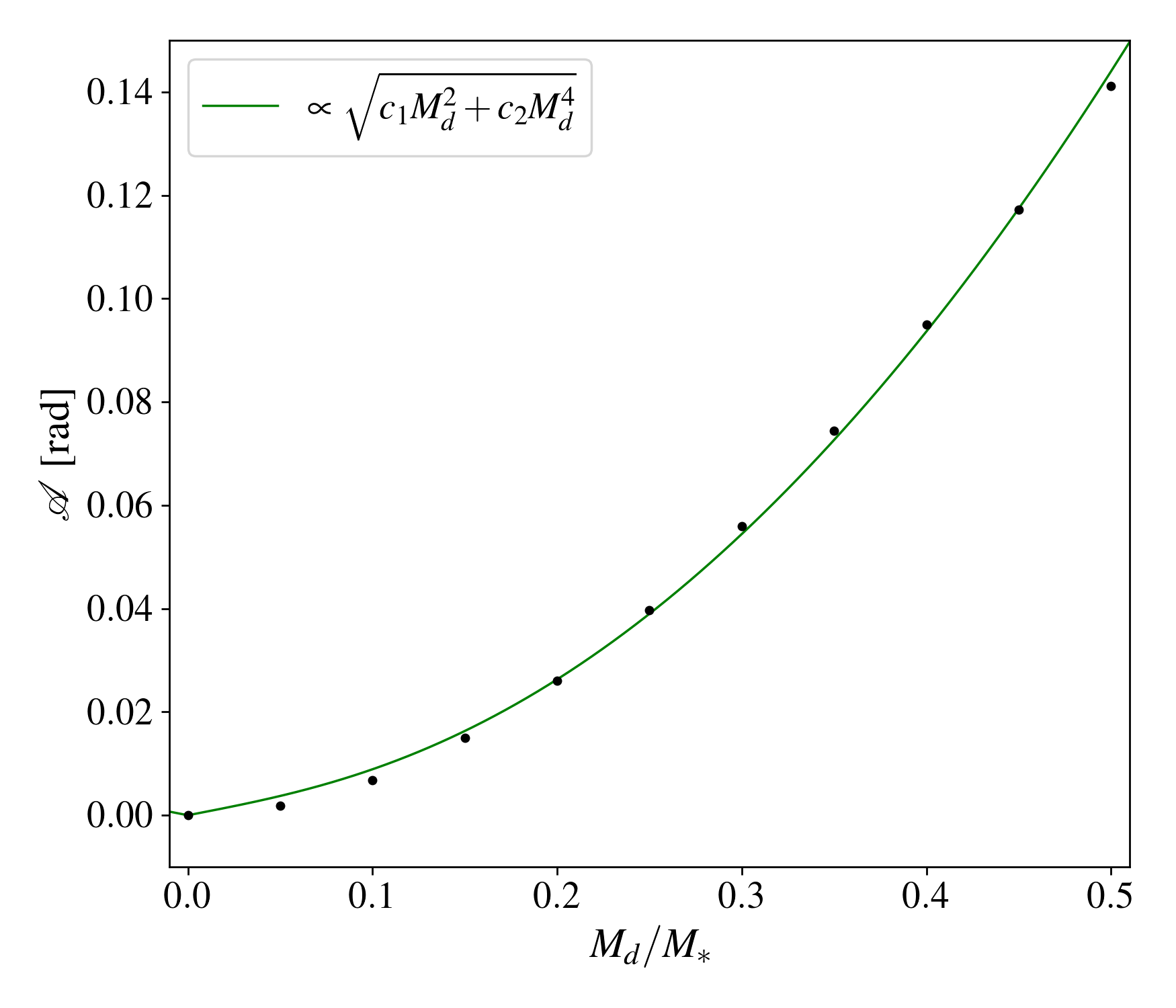}{0.325\textwidth}{(c)} }
    \caption{Schematic view of how the amplitude of the wiggle is computed (a): the red line is the Keplerian channel map, while the blue line is the perturbed one. The amplitude of the perturbation is computed using equation \ref{amplitude}.  Amplitude of the central channel as a function (b) of the cooling parameter $\beta$  {for a disk mass $M_d/M_\star = 0.3$} and (c) of the mass of the disc  {for a cooling parameter
    $\beta = 5$}. The black dots are the results of the analytical model so far described.}
    \label{ampl}
\end{figure}

 {It is possible to quantify the amplitude of the VP considering the integrated geometrical distance between the perturbed and the unperturbed channel. Mathematically, a channel map $C_v$ is a 2D curve defined parametrically from a one-dimensional interval $\mathcal{I}$ to a two-dimensional space $\mathcal{R}^2$. In our case, the two-dimensional space is the cylindrical space $(r,\phi)$ and the interval $\mathcal{I}$ depends on the parameterization we choose: for simplicity, we take $\mathcal{I}=[0,1]$. For a given channel velocity $v$, we call the Keplerian channel map $C_v^k$ and the perturbed one $C_v^p$: mathematically speaking, the two channel maps can be written parametrically as 
\begin{equation}
    C_v^k(s) = \left(\begin{smallmatrix}
    f_{r}^k(s) \\  \\
    f_{\phi}^k(s) \\
    \end{smallmatrix}\right), \quad 
    C_v^p(s) = \left(\begin{smallmatrix}
    f_{r}^p(s) \\  \\
    f_{\phi}^p(s) \\
    \end{smallmatrix}\right),
\end{equation}
where $s$ is a parameterer in the interval $\mathcal{I}$, in our case $s\in[0,1]$. The amplitude of the perturbation is then computed as
\begin{equation}\label{amplitude}
    \mathcal{A}_v = \left[\int_0^1\text{d}s ||C_v^p-C_v^k||^2\right]^{1/2},
\end{equation}
that is the length of the curve $C_v^p - C_v^k$.}
Panel (a) of figure \ref{ampl} schematically shows the quantities involved in equation (\ref{amplitude}). The amplitude is determined by both the cooling factor and the disc mass: a smaller $\beta$ generates a bigger deviation from the Keplerian case, because the amplitude of the density perturbations is inversely proportional to $\beta$ (equation \ref{cool}). On panel (b) of figure \ref{ampl} we show the amplitude of the wiggle ($v_\text{obs} = 0$) as a function of the cooling parameter $\beta$: it is possible to describe the relation between $\mathcal{A}$ and $\beta$ as a power law, with an index of $-1/2$, and this can be easily seen from equations (\ref{vpsdef}). The role of the cooling factor acts as the planet mass in the case of planetary kinks: indeed, the amplitude of the kink is determined by the mass of the embedded protoplanet, and it follows the relation $\mathcal{A}\propto M_p^{1/2}$ \citep{planetkinks}. In addition, the amplitude of the wiggle is also determined by the disc mass: in particular, it affects the perturbed velocities because it is related to the sound speed. Indeed, with the hypothesis of $Q\simeq1$ we get $c_s =u_k {H}/{r} \simeq u_k {M_d}/{M_\star}$.
The disc mass is directly proportional to the sound speed, and then bigger $c_s$  means faster propagation of density waves. Panel (c) of figure \ref{ampl} shows how the amplitude of the wiggle depends on the disc mass. The trend is easily explained by looking at the equations of the VPs (\ref{vpsdef}): in the radial perturbation, the disc mass appears with a quadratic dependence, while in the azimuthal one, it appears linearly. The amplitude of the channel (\ref{amplitude}) is proportional to the root sum of squares, thus $\mathcal{A}\propto\sqrt{c_1M_d^2 + c_2M_d^4}$, where $c_1, c_2$ are two constants that depend on the other parameters, as the cooling.

Thanks to the analytical expression for the VPs, we can infer in what regimes the radial or azimuthal perturbation dominates the wiggle. As we have already noted, for larger $M_d/M_\star$ the radial perturbation dominates. This is a crucial point because a perturbation of the channel $v_\text{obs}=v_\text{syst}$ is visible only if there is a radial perturbation: indeed, as in the Keplerian case, if the velocity is purely azimuthal, the central channel is a straight line.

Unfortunately, the mass of the disc and the cooling $\beta$ are degenerate parameters when we consider the shape of the wiggle: as a matter of fact, they both contribute to its amplitude. Is it possible to break the degeneracy? This can be done if an independent method to measure $M_d$ is available. For example, using again the gas kinematics, one could measure deviations from Keplerianity in the azimuthally averaged rotation curve, and compute the mass of the disc through equation (\ref{discvel}), breaking the degeneracy. \citet{bennielias} did it for the system Elias 2-27, giving a dynamical estimate of the mass of the disc $M_d \simeq 0.08M_\odot \simeq 0.17M_\star$. Conversely, when an approach like this is not possible, we can give a rough estimate of the disc mass through dust emission. Indeed, from dust thermal emission it is possible to measure the dust mass of the disc and then, assuming a dust to gas ratio, we can estimate also the total mass of the disc. 

 {Interestingly, if we look at the perturbed velocity field, there could also be a purely kinematical way to break the degeneracy between disk mass and cooling. Indeed if we write the observed velocity field for $\phi = \pi/2$ (semi-minor axis of the disk) we get 
\begin{equation}
    v_\text{obs} = u_r \sin \theta = \text{Re}\left[\delta u_r(r)e^{i(m\pi/2 + \psi)}\right] \sin \theta = f_1(r) \beta^{-1/2}\left(\frac{M_d}{M_\star}\right)^2 u_k,
\end{equation}
where $f_1$ is a known function of radius; conversely, for $\phi = 0$ (semi-major axis of the disk) we get
\begin{equation}
    v_\text{obs} = u_\phi \sin \theta = \left(r \Omega + \text{Re}\left[\delta u_\phi(r)e^{i\psi}\right]\right)\sin \theta = f_2(r)\beta^{-1/2}\frac{M_d}{M_\star}u_k + r\Omega\sin \theta,
\end{equation}
where $f_2$ is a known function of the radius. Since the perturbed velocities scale differently with disk mass, if we could measure accurately the ratio of the two components of the perturbed velocity, we could in principle directly obtain a measurement of the disk mass. However, we note that it is challenging to extract these information from an actual observation.}

Breaking the degeneracy allows us to constrain the cooling parameter $\beta$, which gives important information about the physical processes that are happening in the protoplanetary environment, and on the tendency of the disc to fragment into bound clumps.

So far we have described what determines the amplitude of the wiggle: as far as its frequency is concerned, it is determined by the pitch angle and by the number of spiral arms. In figure \ref{wiggA} we show the shape of the wiggle for different values of $\alpha_p$ and $m$. We clearly see that the frequency of the wiggle is bigger when decreasing $\alpha_p$ and increasing $m$.

\subsection{Limitations}
As can be seen in figure \ref{wiggA}, the number of spiral arms slightly influences the amplitude of the wiggle: all the calculations have been made under the WKB assumption, that requires $m/rk <<1$. Thus, for high $m$, this assumption is not valid anymore: in fact, we are not considering the relations between the number of spiral arms and the mass of the disc, or its thickness; the only way to take into account these non-linear effects is by means of numerical simulations \citep{terryprep}. This argument is better understood when looking at equation \ref{disp}: in the tight winding approximation (WKB), $m$ does not enter explicitly in the dispersion relation (except in the Doppler-shifted frequency), thus both axisymmetric ($m=0$) and non-axisymmetric ($m\neq0$) perturbations have the same instability threshold. However, it is well known \citep{OPB} that massive discs are subject to large scale non-axisymmetric instabilities even though $Q>1$. Indeed, a local dispersion relation can also be obtained in the case of open spiral structures \citep{LAUB}: it is a cubic rather than quadratic expression in $k$ and depends on a new dimensionless parameter
\begin{equation}
    \mathcal{J} = m \frac{\pi G \Sigma}{r \kappa^{2}} \frac{4 \Omega}{\kappa}\left|\frac{\mathrm{d} \ln \Omega}{\mathrm{d} \ln r}\right|^{1 / 2} \approx \sqrt{6} m \frac{M_{\mathrm{d}}}{M_{*}}.
\end{equation}
Being a function of $m$ and $M_d$, the $ \mathcal{J}$ criterion takes into account how massive the disc is and the number of spiral arms: massive discs are prone to exhibit low-$m$ modes instability. Indeed, there is a link between the number of spiral arms and the mass of the disc \citep{dipierro14}.

Another important point to stress is that we constructed a 2D model of the disc, neglecting its height. Thus, we are basically considering only what happens in the disc mid-plane. The main effect of the disc thickness is to ``dilute" the gravity field, and this can be incorporated into the quadratic dispersion relation
\begin{equation}
    D_\text{thick}(\omega, k, m) = (\omega -m\Omega)^2 -c_s^2 k^2 -\kappa^2 + 2\pi G \Sigma \frac{|k|}{1+|k|H}.
\end{equation}
In this case, the stability criterion only changes slightly, becoming $Q\gtrsim 0.647$ \citep{KrattLod}. Observationally speaking, the disc thickness is important when we take into account the molecular line emission of CO isotopologues. As a matter of fact, $^{12}$CO, that is the most abundant isotopologue, is not a good tracer of the disc mid-plane, because it becomes optically thick at the disc surface. On the contrary, other less abundant CO isotopologues as $^{13}$CO or C$^{18}$O have more optically thin lines and as a consequence they trace better the disc mid-plane \citep{miotello}. 

 {In addition, our analysis takes into account a single spiral mode $m$, while it is well known that for small disk-to-star mass ratio, there could be a superposition of modes. However this is not an actual problem: indeed, we know that after filtering out through the ALMA response (that we do not do in this paper), as shown in \citet{dipierroalma}, only the dominant mode appears. This makes our single-mode analysis still valid.}

\subsection{Constraining the cooling factor - mock test}\label{mock}
So far, we have seen that the cooling factor deeply influences the shape of the channel maps. Thanks to this property, we propose a method with which we can constrain effectively the cooling factor of observed systems. 
In order to verify the accuracy of the calculations we have made, we apply the method just described to a numerical simulation. We perform an SPH simulation using the code PHANTOM \citep{phantom}. This code is widely used in astrophysical community to study gas and dust dynamics in accretion discs \citep{claphantom,enriphnt,planet3}; in this work, we used the so-called ``one fluid" method and we simulated a gas only disc, neglecting the dust component. The initial conditions of the disc are  $r_\text{in} = {1}\text{au}$, $r_\text{out}={50}\text{au}$, $\Sigma\propto r^{-1}$, $M_d = 0.5\text{M}_\star$, $M_\star = {1}{\text{M}_\odot}$. The cooling factor $\beta$ has been set to $\beta=8$ and the two parameters that control the artificial viscosity to $\alpha^{AV} = 0.1$, $\beta^{AV} = 0.2$, in order to reduce as much as possible the effects of artificial dissipation \citep{transport1}.  {The initial sound speed profile follows a simple power law $c_s\propto r^{-0.25}$. However, this profile is rapidly modified by the cooling}. For this simulation, we used $N=5\times10^5$ particles of gas. 
%Panel (a) in figure \ref{sim} shows a snapshot of the simulation after eight outer dynamical times (one thermal time): 
the simulation shows a predominance of the $m=2$ azimuthal mode and the computed pitch angle is $\alpha_p \sim 13^\circ$.

%\begin{figure}
% \gridline{\fig{images/snapshot.png}{0.35\linewidth}{(a)}
 %\fig{images/rotcurve.png}{0.35\linewidth}{(b)}}
 %\caption{Snapshot of the simulation at $t=t_\text{th}=8t_\text{dyn}$ (a), best-fit model (black solid line) for the simulated rotation curve (red markers) with $M_d = 0.5M_\star$ (b). %and comparison of the wiggle between the simulation (green line) and the analytical model (black line)  {(c)}.
 %}
 %\label{sim}
%\end{figure}

\begin{figure*}
	\gridline{\fig{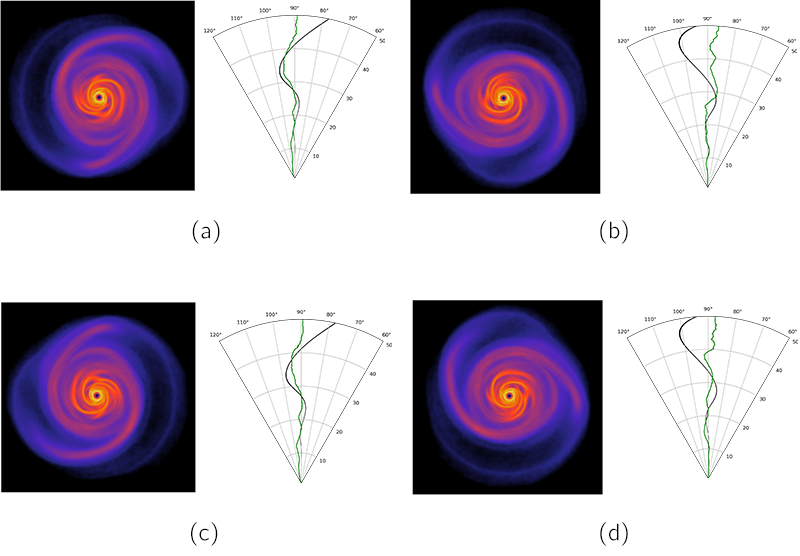}{0.9\linewidth}{}}
    \caption{Comparison between analyical (black line) and numerical (green line) perturbation for different viewing angle of the simulated GI disk. Each picture is rotated along the $z$-axis of an angle of $\pi/2$.}
    \label{comparisons_angle}
\end{figure*}

We now constrain the cooling factor using the method described previously, and make a comparison with the actual value set in the simulation. First of all we compute the rotation curve, azimuthally averaging $u_\phi(r,\phi)$. Then, we find the value of the disc mass that best describe the curve using equation (\ref{discvel}): the best value corresponds to $M_d = 0.5\text{M}_\star$, as expected. 
%In panel (b) of figure \ref{sim} we show the comparison between the simulation data and the best fit rotation curve, where the errors are computed as the azimuthal standard deviation. 
We have broken the degeneracy between the mass and the cooling: thus, we are now able to constrain the cooling $\beta$ through the wiggle amplitude.  {Figure \ref{comparisons_angle} shows the central channel of the projected velocity field of the simulation compared with the one from the analytical model, for different spiral angles: as expected, a wiggle is present. The amplitude of the simulated wiggle\footnote{Here, we refer to panel (a) of Figure \ref{comparisons_angle}.} }is $\mathcal{A}_\text{sim} = {0.11}\text{rad}$. The amplitude of the wiggle as a function of the cooling factor, for the parameters previously reported, is described by $\mathcal{A}(\beta) = \mathcal{A}_\text{in}\beta^{-0.5}$, where $\mathcal{A}_\text{in} = {0.33}\text{rad}$. The estimated cooling factor is then $\beta_\text{sim} = (\mathcal{A}_\text{sim}/\mathcal{A}_\text{in})^{-2} \simeq 9$, that is in good agreement with the real value $\beta = 8$.  { The overestimated value of $\beta$ can actually be interpreted by the lack of viscosity in our analytical model. Indeed, because of its dissipative nature, we expect it to damp GI-driven perturbations, resulting in a less pronounced wiggle. This behaviour is visible in the comparison with the numerical simulations, hence the simulated perturbation appears less wide than the analytical one.}

 {
Figure \ref{comparisons_angle} shows a comparison between the analytical and the numerical wiggle, for different viewing angle of the spiral structure (the observer is assumed to be along a vertical line on the bottom of the images). There is a very good agreement in panel (a) and (c) for which the line of sight intercept the largest extent of the spiral, while the perturbation is overestimated in panel (b) and (d) in which the prominent spiral structure lies on a line perpendicular to the line of sight. For the former case, the agreement between our model and the simulation is remarkable. For the two other orientations, while there is good agreement in the inner disc (where the perturbation is however smaller), our model overestimates the perturbation in the outer disc, where in the simulation the spiral structure vanishes. This suggests that in actual observations, our analysis is most reliable when a density spiral (for example traced by the dust continuum) is also visible superimposed to the kinematical wiggle.
}

\subsection{Conclusions}
In this work we have analytically studied the velocity perturbations in a self-gravitating disc caused by the presence of a spiral density wave in the WKB regime. We then applied this result to obtain the projected velocity field (moment one equivalent) and the channel maps, studying their deviations from the Keplerian case. We found what \citetalias{Hall2020} have already seen from numerical simulations, that deviations from Keplerian rotation are a global phenomenon, resulting in velocity ``kinks" across the entire radial and azimuthal extent of the disc. The kinematics deviations, called GI wiggles, depend on the structure of the spiral density wave, namely its amplitude (connected to the cooling and to the disc mass) and its radial frequency (connected to the pitch angle and to the azimuthal wavenumber). 

\cite{dsharpdev} found nine deviations from Keplerian rotation pattern in the DSHARP circumstellar discs: three of them, Elias 2-27 \citep{eliasperez,teresa}, IM Lup and WaOph6 show also spiral structures in the millimetric continuum emission. These three systems are believed to be self-gravitating \citep{huang}, thus their deviations from Keplerian rotation may be interpreted as wiggles. In addition, \cite{bennielias} have shown that the rotation curve of Elias 2-27 is better described adding the contribution of the disc gravitational potential, meaning that the effects of disc self gravity are not negligible.

Higher resolution observations of systems like those will make it possible to investigate the cooling of protoplanetary discs: indeed, the degeneracy between mass and cooling can be broken by means of the rotation curve, and thus the cooling parameter $\beta$ can be constrained effectively through the wiggle's amplitude, as we have shown in section \ref{mock}. Knowing more about the cooling will give us insights about the gravitational instability process.

\section*{Acknowledgements}
The authors thank the anonymous referee for insightful comments and suggestions, and Richard Booth, Cathie Clarke and Pietro Curone for useful discussions. This project has received funding from the European Union's Horizon 2020 research and innovation programme under the Marie Skłodowska-Curie grant agreement No 823823 (DUSTBUSTERS RISE project). BV acknowledges funding from the ERC CoG project PODCAST No 864965.

%% To help institutions obtain information on the effectiveness of their 
%% telescopes the AAS Journals has created a group of keywords for telescope 
%% facilities.
%
%% Following the acknowledgments section, use the following syntax and the
%% \facility{} or \facilities{} macros to list the keywords of facilities used 
%% in the research for the paper.  Each keyword is check against the master 
%% list during copy editing.  Individual instruments can be provided in 
%% parentheses, after the keyword, but they are not verified.

\vspace{5mm}
\facilities{SPLASH: an interactive visualisation tool for SPH data \citep{splash}}

%% Similar to \facility{}, there is the optional \software command to allow 
%% authors a place to specify which programs were used during the creation of 
%% the manuscript. Authors should list each code and include either a
%% citation or url to the code inside ()s when available.

%% Appendix material should be preceded with a single \appendix command.
%% There should be a \section command for each appendix. Mark appendix
%% subsections with the same markup you use in the main body of the paper.

%% Each Appendix (indicated with \section) will be lettered A, B, C, etc.
%% The equation counter will reset when it encounters the \appendix
%% command and will number appendix equations (A1), (A2), etc. The
%% Figure and Table counter will not reset.

%\appendix

%% For this sample we use BibTeX plus aasjournals.bst to generate the
%% the bibliography. The sample631.bib file was populated from ADS. To
%% get the citations to show in the compiled file do the following:
%%
%% pdflatex sample631.tex
%% bibtext sample631
%% pdflatex sample631.tex
%% pdflatex sample631.tex

\bibliography{sample631}{}
\bibliographystyle{aasjournal}

%% This command is needed to show the entire author+affiliation list when
%% the collaboration and author truncation commands are used.  It has to
%% go at the end of the manuscript.
%\allauthors

%% Include this line if you are using the \added, \replaced, \deleted
%% commands to see a summary list of all changes at the end of the article.
%\listofchanges

\end{document}